# Building Materials Genome from Ground-State Configuration to Engineering Advance

*Zi-Kui Liu*

Department of Materials Science and Engineering, The Pennsylvania State University, University Park, Pennsylvania 16802, USA

E-mail: prof.zikui.liu@psu.edu

Abstract

Individual phases are commonly considered as the building blocks of materials. However, the accurate theoretical prediction of properties of individual phases remains elusive. The top-down approach by decoding genomic building blocks of individual phases from experimental observations is non-unique. The density functional theory (DFT), as the state-of-the-art solution of quantum mechanics, prescribes the existence of a ground-state configuration at zero K for a given system. It is self-evident that the ground-state configuration alone is insufficient to describe a phase at finite temperatures as symmetry-breaking non-ground-state configurations are excited statistically at temperatures above zero K. Our multi-scale entropy approach (recently terms as Zentropy theory) postulates that the entropy of a phase is composed of the sum of the entropy of each configuration weighted by its probability plus the configurational entropy among all configurations. Consequently, the partition function of each configuration in statistical mechanics needs to be evaluated by its free energy rather than total energy. The combination of the ground- and non-ground-state configurations represents the building blocks of materials and can quantitively predict free energy of individual phases with the free energies of ground- and non-ground-state configurations predicted from DFT, plus all properties derived from free energy of individual phases.

# 1. Introduction

Inspired by the success of Human Genome Project (HGP) [1], CALPHAD (CALculation of PHAse Diagrams) modeling method [2], and CALPHAD-based systems materials design [3], the author of the present paper introduced the term "Materials Genome"® in 2002 [4] when incorporating his company Materials Genome, Inc. [5], shortly after he joined the Pennsylvania State University and became the Editor-In-Chief of the CALPHAD Journal [6]. At the same time the author started the integration of first-principles calculations based on density functional theory (DFT), the CALPHAD modeling, phase field simulations, and finite element analysis [7]. The author agreed for the use of Materials Genome by the US government [8] for the Materials Genome Initiative (MGI) [9] and finally trademarked the term [5].

In the CALPHAD method, the models of Gibbs energies of individual phases are developed with model parameters evaluated from computational and experimental thermodynamic data. The CALPHAD modeling starts from pure elements to binary and ternary systems, which are then extrapolated to multicomponent systems, and individual phases were thus considered by the author as the building blocks of materials [10] with their free energies [11] and diffusivity [12] predicted from DFT-based calculations using the ground-state configuration [13]. The impact of the CALPHAD approach was reviewed in a recent special issue that collects the lectures given at the CALPHAD Global 2021 Virtual conference [14–19]. The CALPHAD method was pioneered by Kaufman as described in the monograph by Kaufman and Bernstein [2]. There are two additional books [20,21] dedicated to the CALPHAD method and other books [22–25] that discussed the method. The CALPHAD community consists of the CALPHAD journal, the CALPHAD annual conference, and a nonprofit foundation (CALPHAD, Inc.) [26].

There are well developed commercial tools for CALPHAD modeling and a wide range of CALPHAD databases for education, research, and industrial applications [27–31]. The author's group started to develop an automated CALPHAD modeling tool, Extensible Self-optimizing Phase Equilibria Infrastructure (ESPEI), a while ago with limited success due to the lack of flexible computational engine [32]. Recently, the group developed an open-source software package, PyCalphad, for thermodynamic calculations [33,34] with capability of inserting new models and used it to develop a complete new ESPEI code [35,36] for optimizing model fitness and model parameters and performing uncertainty and sensitivity analysis [37,38].

It is evident that the author's "Materials Genome"® was originally used to denote the individual phases as the building blocks of materials in accordance with the CALPHAD method. The input data for CALPHAD method includes thermochemical data and phase equilibrium data. In principles, thermodynamic modeling could be performed with only thermochemical data as they are the derivatives of free energy, and their integrations give the free energy of each phase. However, most of thermochemical data are derived from measurements of heat that are with large uncertainty, and Gibbs energies of individual phases are thus not accurate enough for accurate prediction of transition temperatures between phases and compositions of phases in equilibrium with each other. Consequently, the Gibbs energy model parameters of all phases need to be refined simultaneously using experimentally measured phase transition data, which limits the predictive power of CALPHAD databases for discovery of new materials.

The natural next step is to seek approaches to accurately predict the free energies of individual phases, so that the phase equilibrium data are not needed as they are not available for new materials to be discovered or designed. In this brief overview, the efforts from the author's group in last fifteen years in searching for building blocks for individual phases are discussed [4]. From the viewpoint of statistical mechanics, an individual phase can be considered as a statistical mixture of various configurations that the phase as a system experiences in accordance with statistical mechanics developed by Gibbs [39]. Starting from the ground-state configuration of a system in terms of DFT at zero K, the system experience non-ground-state configurations at temperatures above zero K. With the free energies of these configurations predicted from the DFT-based calculations, it is shown that the phase transitions and related property anomalies can be calculated accurately, showing remarkable agreement with experimental observations for magnetic [40–42] and ferroelectric materials [43], including singularity at critical points. Consequently, those configurations can be defined as the building blocks and used to predict unknown properties of individual phases in terms of the statistical mechanics. Furthermore, the free energies of individual configurations should be used in calculations of their partition functions since those configurations are not pure quantum configurations. The key challenge in this approach is to sample all configurations that the phase experiences under the experimental conditions.

## 2. Effective Hamiltonian approaches in the literature

DFT [44,45] is the state-of-the-art solution of the multi-body Schrödinger equation [46,47] with several approximations [13,18]. It represents the outcome of the many-body interactions involving both the nuclei and the electrons by a set of one-electron Schrödinger's equations, one for each valence electron. It articulates that there is a ground-state configuration at zero K in each system with the lowest energy that is defined by a unique electron density. The first set of internal degree of freedom (DOF) of the system is the deviation of electron density away from the ground-state electron density as discussed by Kohn and Sham [45] in calculating the free energy of the system using the Mermin formalism. The second set of internal DOF is the phonons due to the displacement of atomic nuclei or lattice vibrations, which can be calculated by either supercell method or linear response theory [11,48,49]. Both electron and phonon DOF preserve the symmetry of the ground-state configuration.

It is important to note that building on the local density approximation (LDA), Perdew and co-workers developed the generalized gradient approximation (GGA) [50–52], in which the exchange-correlation energy is treated as a function of both the local electron density and its gradient, resulting in more accurate predictions of electronic structure and the energy of the ground-state configuration. Their latest strongly constrained and appropriately normed (SCAN) meta-GGA [53,54] with quantitatively correct ground-state results considers the symmetry breaking for some systems regarded as strongly correlated [55,56].

At finite temperatures, a system experiences both symmetry-preserving and symmetry-breaking configurations as stipulated by statistical mechanics [39], while an experimental measurement samples the response of the statistical mixture of all configurations with respect to external stimuli. The common approaches to treat internal symmetry-breaking DOF in the literature are to construct an effective Hamiltonian (eH) and evaluate the model parameters by fitting to DFT-based calculations of the ground-state configurations and some selected symmetry-breaking configurations followed by Monte Carlo (MC) or molecular dynamic (MD) simulations to sample the statistical mixtures of configurations and average their properties.

These approaches inevitably introduce errors in both the selection of eH formalisms and the truncation and the fitting parameters of eH. In most cases, microscopic DOF is evaluated for one given configuration in terms of local occupation with the Ising model, magnetic spins with the Heisenberg model, or the electric dipoles with the Landau theory. Coupling between different types of DOFs is not included automatically, but accommodated as additional, specialized terms [42]. Furthermore, each snapshot in MC/MD simulations representants one statistical mixture of ground-state and non-ground-state configurations under given external constraints. It is possible that not all statistical mixtures of configurations are sampled due to the limited simulation time scale. These limitations prevent quantitative predictions in comparison with experimental observations. There are approaches that directly couple DFT with MD and MC such as *ab initio* molecular dynamics (AIMD) [57–59] and quantum Monte Carlo (QMC) [60–64] with reduced errors, but more limited simulation time scale.

**3. Zentropy theory: Coarse graining of entropy as a quantitative predictive approach**

To improve the agreement between theoretical predictions and experimental observations, the author's team developed a multiscale entropy approach (recently termed as the zentropy theory) that considers both ground-sate and non-ground-state configurations of a system [65–67]. Similar to discrete pure quantum configurations in quantum statistical mechanics [68], the zentropy theory considers that a phase is statistically composed of discrete ground-state and non-ground-state configurations which are not pure quantum configurations. There are two key features of the zentropy theory. The first feature is that the free energies of all configurations are predicted from the DFT-based first-principles calculations. This feature is necessary to include the quantum contributions to entropy, which represent the intrinsic properties of each configuration. Another important aspect of the first feature is its capability to include coherently DOFs related to thermal electronic distribution, phonon vibration, local occupation, magnetic spin, and electric dipole through internal DOFs of individual configurations. The main challenge here is to the limitation on the supercell size in DFT-based calculations due to the constraint of current computing power. A minor challenge is the ergodicity of configurations. Both need to be systematically tested so for the predicted free energy of the phase is converged.

The second feature of the zentropy theory is to use the free energy of each configuration in calculating its partition function instead of total energy in statistical mechanics derived by Gibbs [39] and commonly used in the literature. This feature enables the complete counting of total entropy of the phase from its quantum scale to the experimental scale and maintain the quantum contributions to all scales. This originates from the fact that the ground-sate and non-ground-state configurations used in DFT-based calculations are not pure quantum configurations, and their entropies are not zero at finite temperatures and must be included in order to be able to accurately evaluating the total entropy of the phase. As it will be shown below, this does not only affect the total entropy of the phase, but also changes the probability of each configuration.

The zentropy theory has been successfully applied to predict phase transitions in magnetic materials in last decade [40,41] and more recently in ferroelectrics [43]. In magnetic materials, the symmetry-breaking non-ground-state configurations can be constructed through spin flipping. Under the consideration of collinear magnetic configurations, the total number of ergodic configurations equals to $2^n = m$ in a supercell with $n$ magnetic atoms. The zentropy theory stipulates that the entropy of the system is the weighted sum of each configuration and the statistical entropy among configurations as follows [18,41,42,65–67,69,70]

$$S = \sum_{k=1}^{m} p^k S^k - k_B \sum_{k=1}^{m} p^k \ln p^k \qquad Eq.\ 1$$

where $p^k$ and $S^k$ are the probability and entropy of configuration $k$. The above equation represents the integration of the bottom-up approach from individual configurations, i.e., the first summation, and the top-down approach among individual configurations, i.e., the second summation, which is schematically shown in Figure 1.

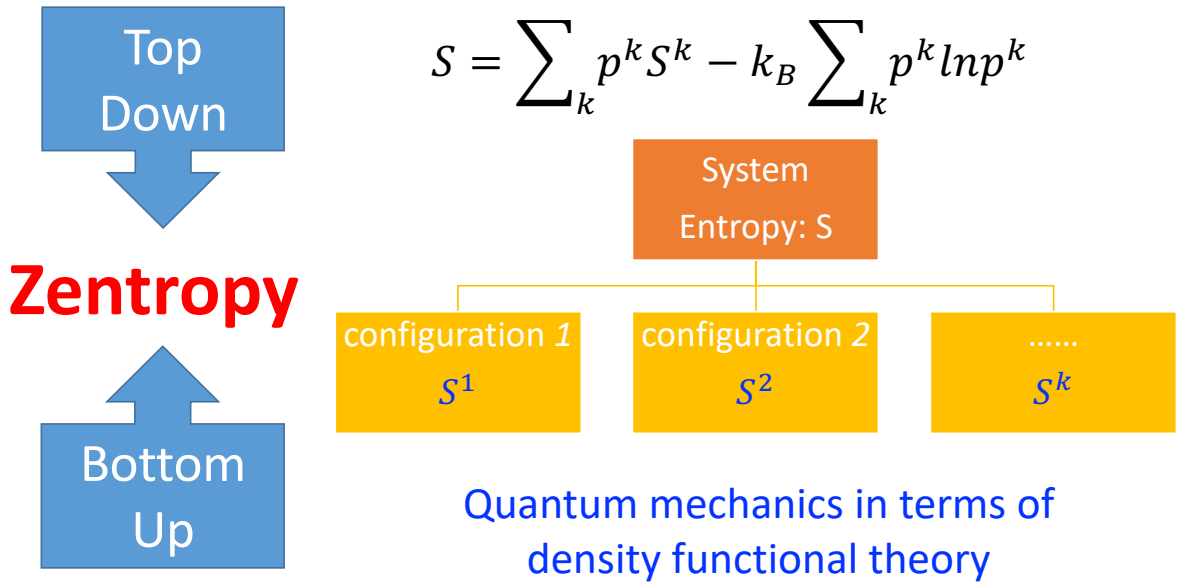

Figure 1: Schematic representation of the zentropy theory with the top-down statistical mechanics that considers the probabilities of independent configurations of the system and bottom-up quantum mechanics with the entropies of ground- and non-ground-state configurations predicted from the density functional theory.

It is noted that the statistical mechanics derived by Gibbs [39] only contains the second summation in Eq. 1, thus only part of the total entropy of the system unless $S^k = 0$ for pure quantum configurations. Furthermore, it is important to point out that Gibbs considered "a great number of independent systems (states) of the same nature (of a system), but differing in the configurations and velocities which they have at a given instant, and differing not merely infinitesimally, but it may be so as to embrace every conceivable combination of configuration and velocities" [39]. He thus broadened the early statistical mechanics from the consideration of the particles of a system to independent systems (configurations of a system), i.e., each configuration of the system must be under the same external conditions as the system. In a canonical ensemble, each configuration and the system thus have the same mass (N), volume (V), and temperature (T), i.e., the same NVT.

Based on Eq. 1, the general formula of statistical mechanics under constant NVT can be written as follows

$$F = \sum_{k=1}^{m} p^k E^k - TS = \sum_{k=1}^{m} p^k F^k - k_B T \sum_{k=1}^{m} p^k \ln p^k \qquad Eq.\ 2$$

$$Z = e^{-\frac{F}{k_B T}} = \sum_{k=1}^{m} Z^k = \sum_{k=1}^{m} e^{-\frac{F^k}{k_B T}} \qquad Eq.\ 3$$

$$p^k = \frac{Z^k}{Z} = e^{-\frac{F^k - F}{k_B T}} \qquad Eq.\ 4$$

where $F$ and $Z$ are the Helmholtz energy and partition function of the system, and $E^k$, $F^k$, and $Z^k$ are the total energy, Helmholtz energy, and partition function of configuration $k$, respectively. The key difference is the use of $F^k$ in Eq. 3 and Eq. 4 instead of $E^k$ as in the Gibbs statistical mechanics [39], which implies $S^k = 0$ for pure quantum state as mentioned above. It is thus self-evident that the properties of all configurations must be evaluated at the same NVT as the system because the statistical combinations of the configurations form the system. It is noted that similar formula was termed as "coarse graining of the partition function" [71–74], though no actual calculations were reported in the literature using the formula by those authors. Similar approaches were also used by Wentzcovitch's group [75,76] as reviewed by the present author [18].

Remarkable agreement between the zentropy-predicted and experimentally observed transition temperatures for a number of magnetic materials has been observed as reviewed previously [40,41] plus the more recent one on $YNiO_3$ with strongly-correlated physics [18,42]. One of unique outcomes of the zentropy theory is the prediction of free energy at unstable states of a system which are between the stable and metastable states of the system, including the critical points predicted in Ce and $Fe_3Pt$ where the system changes from stable to unstable states resulting in bifurcation of the system into two inhomogeneous subsystems [40]. This represents the extreme of anharmonicity in a system that is usually represented by the deviation of entropy or heat capacity away from quasiharmonic behavior [77]. From Eq. 1, it can be seen that the first summation is the linear combination of entropies of individual configurations, and the emergent behaviors, i.e., the behaviors that none of the individual configurations possess, originate from the second summation in the equation.

For ferroelectric (FE) materials with spontaneous electric polarization, the definition of configurations was more challenging due to the strong dipole-dipole interactions that prevent many configurations through simple enumeration of all polarization directions. We explored the configurations of ferroelectric materials using $PbTiO_3$ by means of *ab initio* molecular dynamic simulations (AIMD) [58]. $PbTiO_3$ is one of the most extensively studied ferroelectric materials with the polarized tetragonal ground-state configuration and the macroscopically non-polarized cubic paraelectric (PE) phase above the transition temperature based on X-ray and neutron diffraction data [78]. On the other hand, both experiments [79,80] and AIMD simulations [58] demonstrated that individual Ti-caged unit cells exhibit polarized tetragonal configuration both below and above the FE-PE transition temperature. By following the trajectories of individual Ti atoms (see Fig. 17 in Ref. [40]) and their motions (see video in Supplementary materials in Ref. [43]) using the experimentally determined macroscopic lattice parameters, it was observed in the AIMD simulations that the polarized tetragonal Ti-caged unit cells switch their polarization directions more frequently with the increase of temperature [58]. This process creates more and more misoriented polarized tetragonal Ti-caged unit cells next to each other, resembling the well-known domain walls (DWs) in ferroelectric materials, but through dynamical switching between different polarization directions due to thermal fluctuations rather than freezing in statically.

DWs in ferroelectric materials are discussed extensively in the literature in terms of experimental and computational investigations [81,82]. Based on the experimental and computational results in the literature, there are two types of DWs for tetragonal $PbTiO_3$, i.e., 90° and 180° DWs as twins on the (101) and (100) planes, respectively [83], resulting in three unique configurations including the one without domain wall. Using the DW energies at zero K predicted by the DFT-based first-principles calculations in the literature [83], the macroscopic FE-PE transition temperature predicted by the zentropy theory shows remarkable agreement with experimental measurements [43]. The author's group is currently calculating the Helmholtz energy of each configuration, anticipating better agreements between predicted and measured values.

## 4. Configuration-based materials genome database

As discussed above, the significance of the statistical mechanics by Gibbs [39] is on the consideration of independent configurations that the system experiences with the same NVT as the system, substantially different from the consideration of individual particles in the system, which is not tractable for real materials systems. This is in analogy to the parable of the blind men and the elephant showing the study of the same complex problem from different perspectives and the importance to integrate their insights together [55], i.e., seeing both a forest (top-down) and the trees in the forest (bottom-up). The key capability needed is thus to see the trees in the context of a forest rather than individual trees only, i.e., the possible configurations and their properties in a system, exactly as Gibbs envisioned when he created the statistical mechanics [39].

Historically, knowledge has been primarily accumulated through observations and experimentations, followed by mechanistic understanding and development of fundamental laws. For complex phenomena, phenomenological and mechanical mathematical models were then established with the model parameters fitted to experimental observations and have been used to predict the macroscopic properties of systems, including materials. As any models are intrinsically incomplete and cannot fully represent the complexity underneath observations in general, it is inevitable that there are many different models that are continuously being improved over time along with more in-depth observations. Quantum mechanics developed in the 1920s [46,47] fundamentally changed our understanding of how nature works, and DFT developed in the 1960s enabled the digitization of quantum mechanics [44,45]. DFT starts from the opposite end of the temperature spectrum, i.e., zero K, and the unique ground-state configuration of a given system at zero K. The current mathematical and computational approaches have enabled the accurate prediction of the ground-state configuration of a system, its electronic structure, and associated properties, and more recently its free energy as a function of volume, temperature, and other internal variables.

The missing piece between the observations with multiple configurations and DFT with the ground-state configuration only is thus the non-ground-state configurations in observations that are not considered in typical DFT calculations. Since those configurations are in principle observable, they are metastable and not unstable, and their properties can thus be predicted by DFT in the same way as those of the ground-state configuration. It is self-evident that the ground-state configuration alone is not sufficient to reproduce the observations that also depend on non-ground-state configurations as stipulated by statistical mechanics.

One beauty of statistical mechanics is that the partition function of the system is a simple summation of the partition functions of independent configurations as shown by Eq. 3. Consequently, the free energy of the system can be easily obtained as follows

$$F = -k_B T \ln Z = -k_B T \ln\left(\sum_{k=1}^{m} Z^k\right) = -k_B T \ln\left(\sum_{k=1}^{m} e^{-\frac{F^k}{k_B T}}\right) \qquad Eq.\ 5$$

The probability of each configuration can also be directly calculated arithmetically from Eq. 4. Since the calculation of $F$ does not involve any minimizations, the Helmholtz energy of the system thus obtained represent the Helmholtz energy landscape of the system as a function of internal and external variables of the system, including apexes and valleys. Under given external constraints, the minimization of the Helmholtz energy with respect to internal variables determines whether the system will be in a single-phase state or a multiple-phase state where the probabilities of configurations in all phases are different from each other, commonly referred as miscibility gap in the literature [22,23]. The point between the single-phase and multi-phase states is defined as the critical point where the macroscopically homogeneous single-phase state loses its stability and becomes unstable with the derivative of a potential to its conjugate molar quantity, i.e., the second derivative of the internal or free energy to the molar quantity, approaching from positive to zero. The physical properties of the system defined by the derivatives of a molar quantity to its conjugate potential, i.e., the inverse of the above stability derivative which is also the second derivative of free energy to the potential, diverge and become positive infinite. However, the divergence of properties between a molar quantity and a non-conjugate potential can either be positive or negative as the stability criteria do not prescribe their signs. The predictions of the critical points in Ce and Fe$_3$Pt showed remarkable agreement with experimental observations, including the positive and negative divergences of thermal expansions, i.e., the derivative of volume to temperature, for Ce and $Fe_3Pt$, respectively, i.e., the anti-INVAR and INVAR phenomena in these two materials [70].

Another significance of the Helmholtz energy landscape is the prediction of free energy of the transition state along the pathways between neighboring states in the system, including the inflection points and the free energy of unstable states between the inflection points. Particularly, the free energy at the apex point representing the free energy barrier of the transition, a critical value for the kinetics of the transition. It is important to point out that each configuration itself is stable, i.e., the derivatives between conjugate variables are positive for each configuration, and it is the statistical competition among all configurations that results in those derivatives of the system becoming zero at the inflection points and negative in the unstable states [18,23,40,41]. It is known that the free energy barrier of transition between two states is related to the interfacial and strain energies between them [84]. In the zentropy theory, the strain energy is taken into account by the requirement that the free energies of all configurations in Eq. 3 are evaluated under the same NVT, while the interfaces are built into individual configurations. For systems with defects such as vacancies, dislocations, grain boundaries or grain, they need to be treated as additional internal DOFs thus independent internal variables of free energy [18]. Such a free energy landscape can be used to predict the transport properties in terms of the theory of cross phenomena newly developed by the present author  as shown in

Table 1 [18,41].

*Table 1: Cross phenomenon coefficients represented by derivatives between potentials, symmetrical due to the Maxwell relations with the off diagonal terms that can be either positive or negative* [18,41].

|  | $T$, Temperature | $\sigma$, Stress | $E$, Electrical field | $\mathcal{H}$, Magnetic field | $\mu_i$, Chemical potential |
|---|---|---|---|---|---|
| $T$ | 1 | $-\dfrac{\partial S}{\partial \varepsilon}$ | $-\dfrac{\partial S}{\partial \theta}$ | $-\dfrac{\partial S}{\partial B}$ | $-\dfrac{\partial S}{\partial c_i}$ Partial entropy |
| $\sigma$ | $\dfrac{\partial \sigma}{\partial T}$ | 1 | $-\dfrac{\partial \varepsilon}{\partial \theta}$ | $-\dfrac{\partial \varepsilon}{\partial B}$ | $-\dfrac{\partial \varepsilon}{\partial c_i}$ Partial strain |
| $E$ | $\dfrac{\partial E}{\partial T}$ | $\dfrac{\partial E}{\partial \sigma}$ | 1 | $-\dfrac{\partial \theta}{\partial B}$ | $-\dfrac{\partial \theta}{\partial c_i}$ Partial electrical displacement |
| $\mathcal{H}$ | $\dfrac{\partial \mathcal{H}}{\partial T}$ | $\dfrac{\partial \mathcal{H}}{\partial \sigma}$ | $\dfrac{\partial \mathcal{H}}{\partial E}$ | 1 | $-\dfrac{\partial B}{\partial c_i}$ Partial magnetic induction |
| $\mu_i$ | $\dfrac{\partial \mu_i}{\partial T}$ Thermodiffusion | $\dfrac{\partial \mu_i}{\partial \sigma}$ Stressmigration | $\dfrac{\partial \mu_i}{\partial E}$ Electromigration | $\dfrac{\partial \mu_i}{\partial \mathcal{H}}$ Magnetomigration | $\dfrac{\partial \mu_i}{\partial \mu_j} = -\dfrac{\partial c_j}{\partial c_i} = \dfrac{\Phi_{ii}}{\Phi_{ji}}$ Crossdiffusion |

Based on the discussions above, the ground-state and non-ground-state configurations of a system are the fundamental building blocks of individual phases in the system, which further form the budling blocks for microstructures of materials. The collection of ground-state and non-ground-state configurations can be considered as the materials genome database for prediction of properties of individual phases and properties of materials.

## 5. Summary and perspectives

The present overview paper articulates that the ground-state configuration and non-ground-state configurations derived from the internal DOFs of the ground-state configuration of a system can be considered as building blocks of individual phases of the system. It is relatively simple to determine the non-ground-state configurations in magnetic and ferroelectric phases as demonstrated in our publications due to their straight-forward internal DOFs. Particularly for ferroelectric materials, the number of DW configurations is relatively few [81]. However, for phases of multiferroics, doped with other elements, or containing defects, the number of configurations can be very large, requiring more efficient approaches to predict their free energies. This is where artificial intelligence (AI) can play a very important role [85,86] so free energies of configurations can be efficiently predicted. Consequently, this configuration-based materials database can be used to accurately predict properties of materials based on the zentropy theory without experimental inputs and enable more efficient discovery of new materials and improvement of existing materials for emergent behaviors. New knowledge and data can thus be created and accumulated theoretically and validated by experiments to ultimately enable the full digitization of both cyber and physical spaces of any systems of interest as the core of $4^{th}$ industry revolution [87].

**Acknowledgements**

The author acknowledges the support from Dorothy Pate Enright Professorship at Penn State, National Science Foundation (FAIN-2229690, CMMI-2226976, CMMI-2050069), Department of Energy (DE-SC0023185, DE-NE0009288, DE-AR0001435, DE-NE0008945), Office of Naval Research (N00014-21-1-2608), Army Research Lab, Air Force Research Office, National Aeronautics and Space Administration, and many industrial companies, with the current grants showing in parenthesis.

**Conflict of Interest**

The authors declare no conflict of interests.

**Data Availability Statement**

No data are used in the present work.

**ORCID**

0000-0003-3346-3696

## References

1. Human Genome Project. (2012). Available at: https://www.genome.gov/10001772/all-about-the--human-genome-project-hgp/.
2. Kaufman, L. & Bernstein, H. *Computer Calculation of Phase Diagrams*. (Academic Press Inc., 1970).
3. Olson, G. B. Computational Design of Hierarchically Structured Materials. *Science* **277,** 1237–1242 (1997).
4. Liu, Z. K. Perspective on Materials Genome®. *Chinese Sci. Bull.* **59,** 1619–1623 (2014).
5. The term of Materials Genome coined in 2002 and trademarked in 2004/2012 (78512752/85271561) by MaterialsGenome, Inc. *http://www.materialsgenome.com* Available at: http://tess2.uspto.gov.
6. CALPHAD Journal. Available at: https://www.sciencedirect.com/journal/calphad/.
7. Liu, Z. K., Chen, L.-Q., Raghavan, P., Du, Q., Sofo, J. O., Langer, S. A. & Wolverton, C. An integrated framework for multi-scale materials simulation and design. *J. Comput. Mater. Des.* **11,** 183–199 (2004).
8. Materials Genome Program. Available at: http://www.mgi.gov.
9. National Science and Technology Council. Materials Genome Initiative for Global Competitiveness. (2011). Available at: https://www.mgi.gov/sites/default/files/documents/materials_genome_initiative-final.pdf.
10. Liu, Z. K., Chen, L.-Q., Spear, K. E. & Pollard, C. An Integrated Education Program on Computational Thermodynamics, Kinetics, and Materials Design. (2003). Available at: https://www.tms.org/pubs/journals/JOM/0312/LiuII/LiuII-0312.html.
11. Wang, Y., Liu, Z. K. & Chen, L.-Q. Thermodynamic properties of Al, Ni, NiAl, and Ni3Al from first-principles calculations. *Acta Mater.* **52,** 2665–2671 (2004).
12. Mantina, M., Wang, Y., Arroyave, R., Chen, L. Q., Liu, Z. K. & Wolverton, C. First-Principles Calculation of Self-Diffusion Coefficients. *Phys. Rev. Lett.* **100,** 215901 (2008).
13. Liu, Z. K. First-Principles calculations and CALPHAD modeling of thermodynamics. *J. Phase Equilibria Diffus.* **30,** 517–534 (2009).
14. Zhong, Y., Otis, R., McCormack, S., Xiong, W. & Liu, Z.-K. Summary report of CALPHAD GLOBAL, 2021. *CALPHAD* **81,** 102527 (2023).
15. Ågren, J. CALPHAD and the materials genome A 10 year anniversary. *CALPHAD* **80,**
15

102532 (2023).

16. Hong, Q.-J., van de Walle, A., Ushakov, S. V. & Navrotsky, A. Integrating computational and experimental thermodynamics of refractory materials at high temperature. *CALPHAD* **79,** 102500 (2022).

17. Spencer, P. J. The origins, growth and current industrial impact of Calphad. *CALPHAD* **79,** 102489 (2022).

18. Liu, Z. K. Thermodynamics and its prediction and CALPHAD modeling: Review, state of the art, and perspectives. *CALPHAD* **82,** 102580 (2023).

19. Olson, G. B. & Liu, Z. K. Genomic materials design: CALculation of PHAse Dynamics. *CALPHAD* **82,** 102590 (2023).

20. Saunders, N. & Miodownik, A. P. *CALPHAD (Calculation of Phase Diagrams): A Comprehensive Guide*. (Pergamon, 1998).

21. Lukas, H. L., Fries, S. G. & Sundman, B. *Computational Thermodynammics: The Calphad method*. (Cambridge University Press, 2007).

22. Hillert, M. *Phase Equilibria, Phase Diagrams and Phase Transformations*. (Cambridge University Press, 2007). doi:10.1017/CBO9780511812781

23. Liu, Z. K. & Wang, Y. *Computational Thermodynamics of Materials*. (Cambridge University Press, 2016). doi:10.1017/CBO9781139018265

24. Pelton, A. D. *Phase diagrams and thermodynamic modeling of solutions*. (Elsevier, 2018).

25. Du, Y., Schmid-Fetzer, R., Wang, J., Liu, S., Wang, J. & Jin, Z. *Computational design of engineering materials : fundamentals and case studies*. (Cambridge University Press, 2023).

26. CALPHAD: Annual conference, Foundation, Journal. Available at: https://calphad.org/.

27. Kaufman, L. Foreword. *CALPHAD* **26,** 141 (2002).

28. Hallstedt, B. & Liu, Z.-K. Software for thermodynamic and kinetic calculation and modelling. *CALPHAD* **33,** 265 (2009).

29. Thermo-Calc Software and Databases. Available at: http://www.thermocalc.com/.

30. CompuTherm Software and Databases. Available at: http://www.computherm.com/.

31. FactSage Software and Databases. Available at: https://www.factsage.com/.

32. Shang, S., Wang, Y. & Liu, Z.-K. ESPEI: Extensible, Self-optimizing Phase Equilibrium Infrastructure for Magnesium Alloys. in *Magnesium Technology 2010* (eds. Agnew, S. R., Neelameggham, N. R., Nyberg, E. A. & Sillekens, W. H.) 617–622 (The Minerals, Metals and Materials Society (TMS), Pittsburgh, PA, 2010).

33. Otis, R. & Liu, Z.-K. pycalphad: CALPHAD-based Computational Thermodynamics in Python. *J. Open Res. Softw.* **5,** 1 (2017).

34. PyCalphad: Python library for computational thermodynamics using the CALPHAD method. *https://pycalphad.org*

35. Bocklund, B., Otis, R., Egorov, A., Obaied, A., Roslyakova, I. & Liu, Z. K. ESPEI for efficient thermodynamic database development, modification, and uncertainty quantification: application to Cu–Mg. *MRS Commun.* **9,** 618–627 (2019).

36. ESPEI: Extensible Self-optimizing Phase Equilibria Infrastructure. *https://espei.org*

37. Paulson, N. H., Bocklund, B. J., Otis, R. A., Liu, Z. K. & Stan, M. Quantified uncertainty in thermodynamic modeling for materials design. *Acta Mater.* **174,** 9–15 (2019).

38. Otis, R., Bocklund, B. & Liu, Z. K. Sensitivity estimation for calculated phase equilibria. *J. Mater. Res.* **36,** 140–150 (2021).

39. Gibbs, J. W. *The collected works of J. Willard Gibbs: Vol. II Statistical Mechanics*. (Yale University Press, Vol. II, 1948).

40. Liu, Z. K. Computational thermodynamics and its applications. *Acta Mater.* **200,** 745–792 (2020).

41. Liu, Z. K. Theory of cross phenomena and their coefficients beyond Onsager theorem. *Mater. Res. Lett.* **10,** 393–439 (2022).

42. Du, J., Malyi, O. I., Shang, S.-L., Wang, Y., Zhao, X.-G., Liu, F., Zunger, A. & Liu, Z.-K. Density functional thermodynamic description of spin, phonon and displacement degrees of freedom in antiferromagnetic-to-paramagnetic phase transition in YNiO3. *Mater. Today Phys.* **27,** 100805 (2022).

43. Liu, Z. K., Shang, S.-L., Du, J. & Wang, Y. Parameter-free prediction of phase transition in PbTiO3 through combination of quantum mechanics and statistical mechanics. *Scr. Mater.* **232,** 115480 (2023).

44. Hohenberg, P. & Kohn, W. Inhomogeneous electron gas. *Phys. Rev. B* **136,** B864–B871 (1964).

45. Kohn, W. & Sham, L. J. Self-Consistent Equations Including Exchange and Correlation Effects. *Phys. Rev.* **140,** A1133–A1138 (1965).

46. Schrödinger, E. An Undulatory Theory of the Mechanics of Atoms and Molecules. *Phys. Rev.* **28,** 1049–1070 (1926).

47. Schrödinger, E. Quantisierung als Eigenwertproblem. *Ann. Phys.* **384,** 361–376 (1926).

48. van de Walle, A., Ceder, G. & Waghmare, U. V. First-principles computation of the vibrational entropy of ordered and disordered Ni3Al. *Phys. Rev. Lett.* **80,** 4911 (1998).

49. Wang, Y., Shang, S., Liu, Z.-K. & Chen, L.-Q. Mixed-space approach for calculation of vibration-induced dipole-dipole interactions. *Phys. Rev. B* **85,** 224303 (2012).

50. Langreth, D. C. & Perdew, J. P. Theory of nonuniform electronic systems. I. Analysis of the gradient approximation and a generalization that works. *Phys. Rev. B* **21,** 5469–5493 (1980).

51. Perdew, J. P., Chevary, J. A., Vosko, S. H., Jackson, K. A., Pederson, M. R., Singh, D. J. & Fiolhais, C. Atoms, molecules, solids, and surfaces: Applications of the generalized gradient approximation for exchange and correlation. *Phys. Rev. B* **46,** 6671–6687 (1992).

52. Perdew, J. P. & Wang, Y. Accurate and simple analytic representation of the electron-gas correlation energy. *Phys. Rev. B* **45,** 13244 (1992).

53. Sun, J., Ruzsinszky, A. & Perdew, J. Strongly Constrained and Appropriately Normed Semilocal Density Functional. *Phys. Rev. Lett.* **115,** 036402 (2015).

54. Furness, J. W., Kaplan, A. D., Ning, J., Perdew, J. P. & Sun, J. Accurate and Numerically Efficient r2SCAN Meta-Generalized Gradient Approximation. *J. Phys. Chem. Lett.* **11,** 8208–8215 (2020).

55. Perdew, J. P., Ruzsinszky, A., Sun, J., Nepal, N. K. & Kaplan, A. D. Interpretations of ground-state symmetry breaking and strong correlation in wavefunction and density functional theories. *Proc. Natl. Acad. Sci. U. S. A.* **118,** e2017850118 (2021).

56. Perdew, J. P., Chowdhury, S. T. U. R., Shahi, C., Kaplan, A. D., Song, D. & Bylaska, E. J. Symmetry Breaking with the SCAN Density Functional Describes Strong Correlation in the Singlet Carbon Dimer. *J. Phys. Chem. A* **127,** 384–389 (2023).

57. Car, R. & Parrinello, M. Unified Approach for Molecular-Dynamics and Density-Functional Theory. *Phys. Rev. Lett.* **55,** 2471–2474 (1985).

58. Fang, H., Wang, Y., Shang, S. & Liu, Z. K. Nature of ferroelectric-paraelectric phase transition and origin of negative thermal expansion in PbTiO3. *Phys. Rev. B* **91,** 024104 (2015).

59. Glensk, A., Grabowski, B., Hickel, T., Neugebauer, J., Neuhaus, J., Hradil, K., Petry, W. & Leitner, M. Phonon Lifetimes throughout the Brillouin Zone at Elevated Temperatures from Experiment and Ab Initio. *Phys. Rev. Lett.* **123,** 235501 (2019).

60. Troyer, M. & Wiese, U.-J. Computational Complexity and Fundamental Limitations to Fermionic Quantum Monte Carlo Simulations. *Phys. Rev. Lett.* **94,** 170201 (2005).

61. Needs, R. J., Towler, M. D., Drummond, N. D. & López Ríos, P. Continuum variational and diffusion quantum Monte Carlo calculations. *J. Phys. Condens. Matter* **22,** 023201 (2010).

62. Carlson, J., Gandolfi, S., Pederiva, F., Pieper, S. C., Schiavilla, R., Schmidt, K. E. & Wiringa, R. B. Quantum Monte Carlo methods for nuclear physics. *Rev. Mod. Phys.* **87,** 1067–1118 (2015).

63. Berg, E., Lederer, S., Schattner, Y. & Trebst, S. Monte Carlo Studies of Quantum Critical Metals. *Annu. Rev. Condens. Matter Phys.* **10,** 63–84 (2019).

64. Mondaini, R., Tarat, S. & Scalettar, R. T. Quantum critical points and the sign problem. *Science* **375,** 418–424 (2022).

65. Wang, Y., Hector, L. G., Zhang, H., Shang, S. L., Chen, L. Q. & Liu, Z. K. Thermodynamics of the Ce γ–α transition: Density-functional study. *Phys. Rev. B* **78,** 104113 (2008).

66. Wang, Y., Hector Jr, L. G., Zhang, H., Shang, S. L., Chen, L. Q. & Liu, Z. K. A thermodynamic framework for a system with itinerant-electron magnetism. *J. Phys. Condens. Matter* **21,** 326003 (2009).

67. Wang, Y., Shang, S. L., Zhang, H., Chen, L.-Q. & Liu, Z.-K. Thermodynamic fluctuations in magnetic states: Fe3Pt as a prototype. *Philos. Mag. Lett.* **90,** 851–859 (2010).

68. Landau, L. D. & Lifshitz, E. M. *Statistical Physics*. (Pergamon Press Ltd., 1970).

69. Liu, Z. K., Wang, Y. & Shang, S. Thermal Expansion Anomaly Regulated by Entropy. *Sci. Rep.* **4,** 7043 (2014).

70. Liu, Z. K., Wang, Y. & Shang, S.-L. Zentropy Theory for Positive and Negative Thermal Expansion. *J. Phase Equilibria Diffus.* **43,** 598–605 (2022).

71. Ceder, G. A derivation of the Ising model for the computation of phase diagrams. *Comput. Mater. Sci.* **1,** 144–150 (1993).

72. Asta, M., McCormack, R. & de Fontaine, D. Theoretical study of alloy phase stability in the Cd-Mg system. *Phys. Rev. B* **48,** 748–766 (1993).

73. van de Walle, A. & Ceder, G. The effect of lattice vibrations on substitutional alloy thermodynamics. *Rev. Mod. Phys.* **74,** 11–45 (2002).

74. van de Walle, A. Methods for First-Principles Alloy Thermodynamics. *JOM* **65,** 1523–1532 (2013).

75. Tsuchiya, T., Wentzcovitch, R. M., Da Silva, C. R. S. & De Gironcoli, S. Spin Transition in magnesiowüstite in earth's lower mantle. *Phys. Rev. Lett.* **96,** 198501

(2006).

76. Umemoto, K., Wentzcovitch, R. M., de Gironcoli, S. & Baroni, S. Order–disorder phase boundary between ice VII and VIII obtained by first principles. *Chem. Phys. Lett.* **499,** 236–240 (2010).

77. Fultz, B. Vibrational thermodynamics of materials. *Prog. Mater. Sci.* **55,** 247–352 (2010).

78. Shirane, G., Hoshino, S. & Suzuki, K. X-ray study of the phase transition in lead titanate. *Phys. Rev.* **80,** 1105–1106 (1950).

79. Sicron, N., Ravel, B., Yacoby, Y., Stern, E. A., Dogan, F. & Rehr, J. J. Nature of the ferroelectric phase transition in PbTiO3. *Phys. Rev. B* **50,** 13168–13180 (1994).

80. Ravel, B., Sicron, N., Yacoby, Y., Stern, E. A., Dogan, F., Rehr, J. J., Slcron, N., Yacoby, Y., Stern, E. A., Dogan, F. & Rehr, J. J. Order-disorder behavior in the phase transition of PbTiO3. *Ferroelectrics* **164,** 265–277 (1995).

81. Marton, P., Rychetsky, I. & Hlinka, J. Domain walls of ferroelectric BaTiO3 within the Ginzburg-Landau-Devonshire phenomenological model. *Phys. Rev. B* **81,** 144125 (2010).

82. Grünebohm, A., Marathe, M., Khachaturyan, R., Schiedung, R., Lupascu, D. C. & Shvartsman, V. V. Interplay of domain structure and phase transitions: theory, experiment and functionality. *J. Phys. Condens. Matter* **34,** 073002 (2022).

83. Meyer, B. & Vanderbilt, D. Ab initio study of ferroelectric domain walls in PbTiO3. *Phys. Rev. B* **65,** 104111 (2002).

84. Porter, D. A., Easterling, K. E. & Sherif, M. *Phase Transformations in Metals and Alloys*. (CRC PRESS, 2021).

85. Krajewski, A. M., Siegel, J. W., Xu, J. & Liu, Z. K. Extensible Structure-Informed Prediction of Formation Energy with improved accuracy and usability employing neural networks. *Comput. Mater. Sci.* **208,** 111254 (2022).

86. Li, K., DeCost, B., Choudhary, K., Greenwood, M. & Hattrick-Simpers, J. A critical examination of robustness and generalizability of machine learning prediction of materials properties. *npj Comput. Mater.* **9,** 55 (2023).

87. Liu, Z. K. Materials 4.0 and the Materials Genome Initiative. *Adv. Mater. Process.* **178(2),** 50 (2020).

Zi-Kui Liu is the Dorothy Pate Enright Professor at Penn State University where he joined in 1999. He obtained his PhD from KTH in Sweden in 1992 and coined the term "Materials Genome®" in 2002. His current research focuses on (1) DFT-based first-principles calculations and deep neural network machine learning for prediction of materials properties in terms of zentropy theory and theory of cross phenomena, (2) their applications for designing materials chemistry, processing, and performances. His team developed the zentropy theory for accurate prediction of free energy of a phase and the theory of cross phenomena for transport properties.

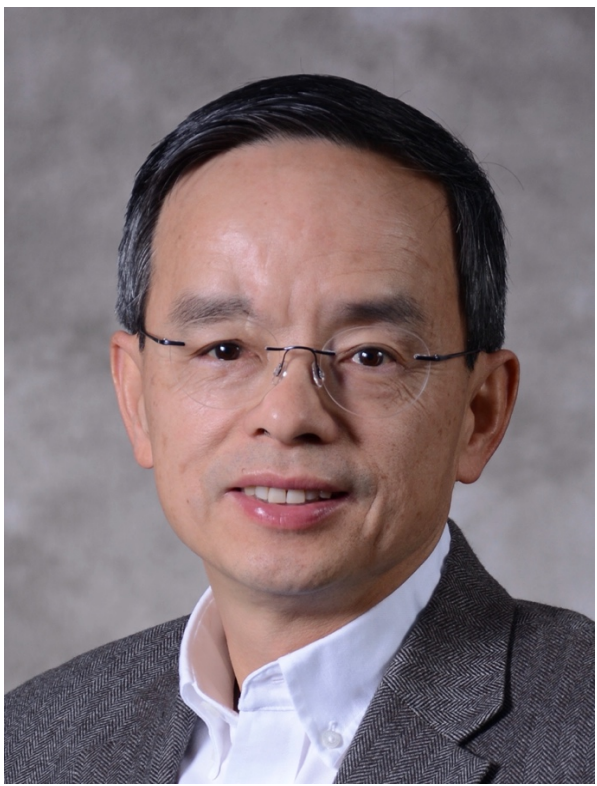